\def\ni{\noindent}
\def\vs{\vskip.3cm}

\def\+{{(+)}}  \def\-{ {(-)} }   \def\0{ {(0)} }
\def\1{ {(1)} }  \def\2{ {(2)} }

\def\m{\mu}

\def\sq{Q\kern-6pt/}
\def\sQ{Q\kern-12pt\nearrow}
\documentclass[11pt,eqs]{article}

\usepackage{latexsym}
\usepackage{amssymb}
\def\be{\begin{equation}}             \def\ee{\end{equation}}
\def\ba{\begin{array}{rcl}}           \def\ea{\end{array}}
\def\beqa{\begin{eqnarray} }          \def\eeqa{\end{eqnarray} }
\def\beqalign{\begin{eqalign}}        \def\eeqalign{\end{eqalign}}
             
\def\bsubeq{\begin{subequations}}     \def\esubeq{\end{subequations}}
\def\bitem{\begin{itemize}}           \def\eitem{\end{itemize}}

\def\DJ{\leavevmode\setbox0=\hbox{D}\kern0pt
 \rlap{\kern.04em\raise.188\ht0\hbox{-}}D}
\def\dj{\leavevmode\setbox0=\hbox{d}\kern0pt
 \rlap{\kern.215em\raise.46\ht0\hbox{-}}d}

\newcommand{\bd}{\begin{displaymath}}
\newcommand{\ed}{\end{displaymath}}
\begin{document}

\title{ Noncommutativity relations in type IIB theory and their supersymmetry
\thanks{Work supported in part by the Serbian Ministry of Science and
Technological Development, under contract No. 141036.}}
\author{B. Nikoli\'c \thanks{e-mail address: bnikolic@ipb.ac.rs} and B. Sazdovi\'c
\thanks{e-mail address: sazdovic@ipb.ac.rs}\\
       {\it Institute of Physics}\\{\it University of Belgrade}\\{\it P.O.Box 57, 11001 Belgrade, Serbia}}
%\date{}
\maketitle
\begin{abstract}

In the present paper we investigate noncommutativity of $D9$
and $D5$-brane world-volumes embedded in space-time of type IIB
superstring theory. Boundary conditions, which preserve half of
the initial supersymmetry, are treated as canonical constraints.
Solving the constraints we obtain original coordinates in terms of
the effective coordinates and momenta. Presence of momenta induces
noncommutativity of string endpoints. We show that
noncommutativity relations are connected by $N=1$ supersymmetry transformations
and noncommutativity parameters are components of $N=1$
supermultiplet.

\end{abstract}
\vs

\ni {\it PACS number(s)\/}: 02.40.Gh, 11.30.Pb, 11.25.Uv, 11.25.-w. 
\par

\section{Introduction}
\setcounter{equation}{0}

In the present paper we investigate the noncommutativity of type
IIB superstring theory  \cite{jopol} in pure spinor formulation
(up to the quadratic terms) \cite{berko} using canonical
approach. We consider two cases: when $D9$-brane is space-time
filling and when $D5$-brane is embedded in space-time. Also we investigate the supersymmetry of noncommutativity relations.

The field content of R-R sector determines stable $Dp$-branes \cite{jopol} in the certain superstring theory. The R-R sector of type IIB theory
contains gauge fields $A_{(0)}$, $A_{(2)}$ and $A_{(4)}$, and
consequently, $Dp$-branes with odd value of $p$ are stable. As a
particular choice, besides $D9$-brane, we will embed $D5$-brane in
ten dimensional space-time.

Spinorial part of ten dimensional superspace is spanned by two
fermionic coordinates, $\theta^\alpha$ and $\bar\theta^\alpha$
$(\alpha=1,2,\dots,16)$, which are Majorana-Weyl spinors. It is
useful to express ten dimensional Majorana-Weyl spinor $S^\alpha$
in terms of two independent $D5$-brane opposite chirality Weyl
spinors, $S^{\alpha_1}$ and $S^{\alpha_2}$
$(\alpha_1,\alpha_2=1,2,\dots,8)$ \cite{jopol,duf,grk}.

In the case of $D9$-brane, when it fills all space-time, we chose
Neumann boundary conditions for all bosonic coordinates $x^\mu$. The boundary condition
for fermionic coordinates,
$(\theta^\alpha-\bar\theta^\alpha)|_0^\pi=0$ produces additional
one for their canonically conjugated momenta,
$(\pi_\alpha-\bar\pi_\alpha)|_0^\pi=0$. Choosing Neumann boundary conditions for
$x^i$ coordinates ($i=0,1,\dots,5$), and Dirichlet boundary
conditions for orthogonal ones $x^a$ ($a=6,\dots,9$) we embed
$D5$-brane in ten dimensional space-time. For fermionic coordinates we choose boundary
condition
$[\theta^\alpha+({}^\star\Gamma\bar\theta)^\alpha]|_0^\pi=0$,
where
${}^\star\Gamma=\Gamma^0\Gamma^1\Gamma^2\Gamma^3\Gamma^4\Gamma^5$
is introduced to preserve supersymmetry \cite{jopol}.
Corresponding boundary condition for momenta is of the form
$[\pi_\alpha+({}^\star\Gamma\bar\pi)_\alpha]|_0^\pi=0$. In terms
of $D5$-brane spinors boundary conditions have the form
$(\theta^{\alpha_1}-\bar\theta^{\alpha_1})|_0^\pi=0$,
$(\theta^{\alpha_2}+\bar\theta^{\alpha_2})|_0^\pi=0$,
$(\pi_{\alpha_1}-\bar\pi_{\alpha_1})|_0^\pi=0$ and
$(\pi_{\alpha_2}+\bar\pi_{\alpha_2})|_0^\pi=0$.

We treat boundary conditions as canonical constraints \cite{kanonski,BNBS,BNBSPLB,BNBSNPB}. Using their
consistency conditions, we rewrite them in compact
$\sigma$-dependent form and find their Poisson brackets. It turns
out that all constraints are of the second class for nonsingular
open string metric $G^{eff}=G-4BG^{-1}B$. Solving the second class
constraints, we obtain initial coordinates in terms of effective
coordinates and momenta. Presence of the momenta in the solutions
for initial coordinates is source of noncommutativity.
Noncommutativity relations are consistent with $N=1$
supersymmetry transformations. We obtained that noncommutativity parameters
contain only odd powers of background fields antisymmetric under
world-sheet parity transformation $\Omega:\sigma\to -\sigma$. They are components of $N=1$ supermultiplet. This result represents a
supersymmetric generalization of the result obtained by Seiberg
and Witten \cite{SW}.

At the end we give some concluding remarks.

\section{Type IIB superstring theory and embedded $D5$-brane}
\setcounter{equation}{0}

We will investigate pure spinor formulation
\cite{berko,susyNC,BNBSPLB,BNBSNPB} of type IIB theory, neglecting ghost
terms and keeping quadratic ones as in the action of
Ref.\cite{susyNC}.

The action in a flat background
\begin{equation}
S_0=\int_\Sigma d^2\xi \left( \frac{\kappa}{2}\eta^{mn}\eta_{\mu\nu}\partial_m x^\mu \partial_n x^\nu-\pi_\alpha \partial_{-} \theta^\alpha+\partial_+ \bar\theta^\alpha \bar\pi_\alpha\right)\, ,
\end{equation}
deformed by integrated form of the massless IIB supergravity
vertex operator
\begin{equation}
V_{SG}=\int_\Sigma d^2 \xi X^T_M A_{MN}\bar X_N\, ,
\end{equation}
produces the full action
\begin{equation}
S=S_0+V_{SG}\, .
\end{equation}

The world sheet ($\Sigma$) parameters are $\xi^m=(\tau\,
,\sigma)$, while $D=10$-dimensional space-time coordinates are labelled by $x^\mu$ ($\mu=0,1,2,\dots,9$). The fermionic extension
of space-time is expressed by same chirality fermionic coordinates
$\theta^\alpha$ and $\bar\theta^{\alpha}$. The variables
$\pi_\alpha$ and $\bar \pi_{\alpha}$ are canonically
conjugated to the coordinates $\theta^\alpha$ and $\bar\theta^{\alpha}$, respectively. The fermionic coordinates and momenta are
Majorana-Weyl spinors.

Using equations of motion which are consequences of BRST invariance, requiring for all background fields to be constant and restricted the action to the quadratic terms, the vertex operator gets the form
\begin{eqnarray}
V_{SG}=\int_\Sigma d^2\xi \left[
\kappa(\frac{1}{2}g_{\mu\nu}+B_{\mu\nu})\partial_+x^\mu\partial_-x^\nu-\pi_\alpha
\Psi^\alpha_\mu \partial_-x^\mu+\partial_+ x^\mu
\bar\Psi^\alpha_\mu\bar\pi_\alpha+\frac{1}{2\kappa}\pi_\alpha
F^{\alpha\beta}\bar\pi_\beta\right]\, ,
\end{eqnarray}
where $g_{\mu\nu}$ is symmetric, $B_{\mu\nu}$ is antisymmetric Neveu-Schwarz
field, $\Psi^\alpha_\mu$ and $\bar\Psi^\alpha_\mu$ are NS-R gravitino fields and $F^{\alpha\beta}$ is R-R field strength. Adding $V_{SG}$ to flat background action, we have
\begin{eqnarray}\label{eq:SB}
&{}&S=\kappa \int_\Sigma d^2\xi \left[
\frac{1}{2}\eta^{mn}G_{\mu\nu}+\varepsilon^{mn}
B_{\mu\nu}\right]\partial_m x^\mu \partial_n x^\nu
\\&+&\int_\Sigma d^2 \xi \left[ -\pi_\alpha
\partial_-(\theta^\alpha+\Psi^\alpha_\mu
x^\mu)+\partial_+(\bar\theta^{\alpha}+\bar \Psi^{\alpha}_\mu
x^\mu)\bar\pi_{\alpha}+\frac{1}{2\kappa}\pi_\alpha F^{\alpha
\beta}\bar \pi_{\beta}\right ]\, ,\nonumber
\end{eqnarray}
where $G_{\mu\nu}=\eta_{\mu\nu}+g_{\mu\nu}$ is constant gravitational field.

Embedding $D9$-brane means that we choose Neumann boundary
conditions for all space-time coordinates $x^\mu$ so that
$D9$-brane fills whole space-time. In order to embed $D5$-brane in
ten dimensional space-time we choose Neumann boundary conditions
for $x^i\, (i=0,1,\dots,5)$ and Dirichlet boundary conditions for
orthogonal directions $x^a\, (a=6,7,8,9)$. The choice of
background fields is the same as in Ref.\cite{BNBSNPB} and the
action is of the form
\begin{eqnarray}\label{eq:SB1}
&S&=\kappa \int_\Sigma d^2\xi \left[
\frac{1}{2}\eta^{mn}G_{ij}+\varepsilon^{mn}
B_{ij}\right]\partial_m x^i \partial_n x^j\nonumber
\\&+&2\Re\left\lbrace \int_\Sigma d^2 \xi \left[  -\pi_{\alpha_1}
(\partial_\tau-\partial_\sigma)\left(
\theta^{\alpha_1}+\Psi^{\alpha_1}_i x^i\right)
+(\partial_\tau+\partial_\sigma)\left( \bar\theta^{\alpha_1}+\bar
\Psi^{\alpha_1}_i x^i\right) \bar\pi_{\alpha_1}\right]\right\rbrace  \nonumber
\\ &+&2\Re\left\lbrace \int_\Sigma d^2 \xi \left[  -\pi_{\alpha_2}
(\partial_\tau-\partial_\sigma)
\left(\theta^{\alpha_2}+\Psi^{\alpha_2}_i x^i\right)
+(\partial_\tau+\partial_\sigma) \left(\bar\theta^{\alpha_2}+\bar\Psi^{\alpha_2}_i x^i\right)\bar\pi_{\alpha_2}\right]\right\rbrace  \nonumber
\\&+&\frac{1}{\kappa}\Re\left\lbrace \int_\Sigma d^2\xi\left[\pi_{\alpha_1} f_{11}^{\alpha_1\beta_1}\bar\pi_{\beta_1}+\pi_{\alpha_1} f_{14}^{\alpha_1\beta_1}\bar\pi_{\beta_1}^*-\pi_{\alpha_2} f_{22}^{\alpha_2\beta_2}\bar\pi_{\beta_2}-\pi_{\alpha_2} f_{23}^{\alpha_2\beta_2}\bar\pi_{\beta_2}^*\right]\right\rbrace \nonumber \\ &+&\frac{1}{\kappa}\Re\left\lbrace \int_\Sigma d^2\xi\left[\pi_{\alpha_2} f_{21}^{\alpha_2\beta_1}\bar\pi_{\beta_1}-\pi_{\alpha_1} f_{12}^{\alpha_1\beta_2}\bar\pi_{\beta_2}+\pi_{\alpha_2} f_{24}^{\alpha_2\beta_1}\bar\pi_{\beta_1}^*-\pi_{\alpha_1} f_{13}^{\alpha_1\beta_2}\bar\pi_{\beta_2}^*\right]\right\rbrace\,
,
\end{eqnarray}
where $\Re$ means real part of some complex number, ${}^*$ means
complex conjugation and with $f_{rs}$ we denoted 8 independent
$D5$-brane components of $F^{\alpha\beta}$ (for more details see
Appendix B of Ref.\cite{BNBSNPB}).

\section{Canonical analysis}
\setcounter{equation}{0}

Here we will perform canonical analysis of type IIB superstring
theory. Boundary conditions will be treated as canonical
constraints. Consistency procedure for boundary conditions enable
us to rewrite them in compact $\sigma$-dependent form. It turns
out that all constraints are of the second class.

\subsection{Hamiltonian}

Using standard
canonical procedure we find canonical Hamiltonian of type IIB
superstring theory in the form
\begin{equation}\label{eq:initialham9}
H_c=\int d\sigma \mathcal H_c\, ,\quad \mathcal H_c=T_--T_+\, ,\quad T_{\pm} =t_{\pm}-\tau_{\pm}\, ,
\end{equation}
where
\begin{eqnarray}
t_{\pm}&=&\mp\frac{1}{4\kappa}G^{\mu\nu}I_{\pm \mu}I_{\pm \nu}\,
,\quad I_{\pm \mu}=\pi_\mu+2\kappa \Pi_{\pm
\mu\nu}x'^\nu+\pi_\alpha \Psi^\alpha_\mu-\bar\Psi^{\alpha}_\mu
\bar\pi_{\alpha}\, ,\nonumber \\
\tau_+&=&(\theta'^\alpha+\Psi^\alpha_\mu x'^\mu)
\pi_\alpha-\frac{1}{4\kappa}\pi_\alpha
F^{\alpha\beta}\bar\pi_{\beta}\, ,\quad
\tau_-=(\bar\theta'^\alpha+\bar\Psi^\alpha_\mu
x'^\mu)\bar\pi_{\alpha}+\frac{1}{4\kappa}\pi_\alpha
F^{\alpha\beta}\bar\pi_{\beta}\, . \label{eq:struja9}
\end{eqnarray}
For the case of embedded $D5$-brane canonical Hamiltonian gets the form
\begin{eqnarray}
t_{\pm}&=&\mp\frac{1}{4\kappa}G^{ij}I_{\pm i}I_{\pm j}\, ,\nonumber \\ I_{\pm i}&=&\pi_i+2\kappa \Pi_{\pm
ij}x'^j+2\Re\left(\pi_{\alpha_1} \Psi^{\alpha_1}_i+\pi_{\alpha_2} \Psi^{\alpha_2}_i-\bar\Psi^{\alpha_1}_i
\bar\pi_{\alpha_1}-\bar\Psi^{\alpha_2}_i
\bar\pi_{\alpha_2}\right),\nonumber \\
\tau_+&=&2\Re\left[ \left(\theta'^{\alpha_1}+\Psi^{\alpha_1}_i x'^i\right)
\pi_{\alpha_1}+\left(\theta'^{\alpha_2}+\Psi^{\alpha_2}_i x'^i\right)\pi_{\alpha_2}\right]\nonumber \\&-&\frac{1}{2\kappa}\Re\left( \pi_{\alpha_1} f_{11}^{\alpha_1\beta_1}\bar\pi_{\beta_1}+\pi_{\alpha_1} f_{14}^{\alpha_1\beta_1}\bar\pi_{\beta_1}^*-\pi_{\alpha_2} f_{22}^{\alpha_2\beta_2}\bar\pi_{\beta_2}-\pi_{\alpha_2} f_{23}^{\alpha_2\beta_2}\bar\pi_{\beta_2}^*\right)\nonumber \\ &-&\frac{1}{2\kappa}\Re\left( \pi_{\alpha_2} f_{21}^{\alpha_2\beta_1}\bar\pi_{\beta_1}-\pi_{\alpha_1} f_{12}^{\alpha_1\beta_2}\bar\pi_{\beta_2}+\pi_{\alpha_2} f_{24}^{\alpha_2\beta_1}\bar\pi_{\beta_1}^*-\pi_{\alpha_1} f_{13}^{\alpha_1\beta_2}\bar\pi_{\beta_2}^*\right)\, ,\nonumber \\
\tau_-&=&2\Re\left[ \left(
\bar\theta'^{\alpha_1}+\bar\Psi^{\alpha_1}_i x'^i\right)
\bar\pi_{\alpha_1}+\left(\bar\theta'^{\alpha_2}+\bar\Psi^{\alpha_2}_i x'^i\right)\bar\pi_{\alpha_2}\right]\nonumber
\\&+&\frac{1}{2\kappa}\Re\left( \pi_{\alpha_1} f_{11}^{\alpha_1\beta_1}\bar\pi_{\beta_1}+\pi_{\alpha_1} f_{14}^{\alpha_1\beta_1}\bar\pi_{\beta_1}^*-\pi_{\alpha_2} f_{22}^{\alpha_2\beta_2}\bar\pi_{\beta_2}-\pi_{\alpha_2} f_{23}^{\alpha_2\beta_2}\bar\pi_{\beta_2}^*\right)\nonumber \\ &+&\frac{1}{2\kappa}\Re\left( \pi_{\alpha_2} f_{21}^{\alpha_2\beta_1}\bar\pi_{\beta_1}-\pi_{\alpha_1} f_{12}^{\alpha_1\beta_2}\bar\pi_{\beta_2}+\pi_{\alpha_2} f_{24}^{\alpha_2\beta_1}\bar\pi_{\beta_1}^*-\pi_{\alpha_1} f_{13}^{\alpha_1\beta_2}\bar\pi_{\beta_2}^*\right) \, ,
\label{eq:struja}
\end{eqnarray}
and $\pi_i$, $\pi_{\alpha_1}$, $\pi_{\alpha_2}$,
$\bar\pi_{\alpha_1}$ and $\bar\pi_{\alpha_2}$ are canonically
conjugated to $x^i$, $\theta^{\alpha_1}$, $\theta^{\alpha_2}$,
$\bar\theta^{\alpha_1}$ and $\bar\theta^{\alpha_2}$, respectively.
Note, that in both cases energy-momentum tensor components
$T_{\pm}$ satisfy Virasoro algebra as a consequence of two
dimensional diffeomorphisms.

\subsection{Boundary conditions as canonical constraints}

As a time translation generator Hamiltonian has to be
differentiable with respect to coordinates and their canonically
conjugated momenta. From this fact, following method of
Ref.\cite{BNBS}, we will derive boundary conditions directly in
terms of the canonical variables. Varying Hamiltonian $H_c$, we
obtain
\begin{equation}
\delta H_c=\delta H_c^{(R)}-[\gamma_\mu^{(0)}\delta x^\mu+
\pi_\alpha\delta\theta^\alpha+\delta
\bar\theta^{\alpha}\bar\pi_{\alpha}]\big |_0^\pi\, ,
\end{equation}
where $\delta H_c^{(R)}$ is regular term without $\tau$ and
$\sigma$ derivatives of supercoordinates and supermomenta
variations and
\begin{equation}
\gamma_{\mu}^{(0)}=\Pi_{+ \mu}{}^\nu I_{- \nu}+\Pi_{- \mu}{}^\nu I_{+
\nu}+\pi_{\alpha}
\Psi^{\alpha}_\mu+\bar\Psi^{\alpha}_\mu \bar\pi_{\alpha}\, .
\end{equation}
Consequently, differentiability of Hamiltonian for type IIB theory
demands
\begin{equation}\label{eq:BC1}
\left[\gamma_\mu^{(0)}\delta x^\mu+
\pi_\alpha\delta\theta^\alpha+\delta
\bar\theta^{\alpha}\bar\pi_{\alpha}\right]\Big |_0^\pi=0\, .
\end{equation}

Embedding $D9$-brane implies Neumann boundary conditions for
$x^\mu$ coordinates, which means
\begin{equation}\label{eq:gamami9}
\gamma_\mu^{(0)}|_0^\pi=0\, .
\end{equation}
Boundary condition for fermionic coordinates chosen to preserve
half of the initial $N=2$ supersymmetry is
\begin{equation}\label{eq:gu1}
(\theta^\alpha-\bar\theta^{\alpha})\Big |_0^\pi=0\, ,
\end{equation}
and it produces additional boundary condition for fermionic momenta
\begin{equation}\label{eq:gu2}
\left(\pi_\alpha-\bar\pi_\alpha\right)|_0^\pi=0\, .
\end{equation}

In order to embed $D5$-brane, for $D5$-brane coordinates $x^i$ we
choose Neumann boundary conditions, implying
\begin{eqnarray}\label{eq:gamami}
&{}&\gamma_i^{(0)}\big |_0^\pi=0\, ,\\ &{}&
\gamma_{i}^{(0)}=\Pi_{+ i}{}^j I_{- j}+\Pi_{- i}{}^j I_{+
j}+2\Re\left( \pi_{\alpha_1} \Psi^{\alpha_1}_i+\pi_{\alpha_2}
\Psi^{\alpha_2}_i+\bar\Psi^{\alpha_1}_i
\bar\pi_{\alpha_1}+\bar\Psi^{\alpha_2}_i \bar\pi_{\alpha_2}\right)
\, .\nonumber
\end{eqnarray}
For othogonal coordinates we choose Dirichlet ones, $\delta
x^a|_0^\pi=0$. As in Ref.\cite{BNBSNPB}, dynamics of $x^a$
directions decouples from the rest part of action and we will not
consider this boundary condition. Fermionic boundary conditions
take the form
\begin{equation}\label{eq:gu}
\left[\theta^\alpha+({}^\star\Gamma
\bar\theta)^\alpha\right]|_0^\pi=0\, ,\quad
\left[\pi_\alpha+({}^\star\Gamma\bar\pi)_\alpha\right]|_0^\pi=0\,
,
\end{equation}
where
${}^\star\Gamma=\Gamma^0\Gamma^1\Gamma^2\Gamma^3\Gamma^4\Gamma^5$.
By convention we introduce ${}^\star\Gamma$ because if $Q_1$ and
$Q_2$ are type IIB supersymmetry charges then, after $Dp$-brane is
embedded, the conserved supersymmetry is the linear combination
\cite{jopol}
\begin{equation}
Q_1+\Gamma^0\Gamma^1\dots\Gamma^p Q_2\, .
\end{equation}

Note that arbitrary Majorana-Weyl spinor can be expressed in terms
of two opposite chirality $D5$-brane Weyl spinors
\begin{equation}\label{eq:obliks}
S^\alpha=\left(
\begin{array}{c}
S^{\alpha_1}\\
S^{\alpha_2}\\
(b_1 S^*)^{\alpha_2}\\
-(b_1 S^*)^{\alpha_1}
\end{array}\right)\, ,
\end{equation}
where $b_1$ is $D5$-brane complex conjugation operator. In terms
of $D5$-brane spinors boundary conditions takes the form
\begin{equation}\label{eq:bcf1}
(\theta^{\alpha_1}-\bar\theta^{\alpha_1})\Big |_0^\pi=0\, ,\quad (\theta^{\alpha_2}+\bar\theta^{\alpha_2})\Big |_0^\pi=0\, ,
\end{equation}
\begin{equation}\label{eq:bcf2}
(\pi_{\alpha_1}-\bar\pi_{\alpha_1})\big |_0^\pi=0\, , \quad (\pi_{\alpha_2}+\bar\pi_{\alpha_2})\big |_0^\pi=0\, .
\end{equation}
According with Ref.\cite{BNBS}, we will treat the
expressions (\ref{eq:gamami9})-(\ref{eq:gu2}) and (\ref{eq:gamami}), (\ref{eq:bcf1}) and (\ref{eq:bcf2}) as canonical
constraints for $D9$ and $D5$-brane, respectively.

\subsection{Consistency of constraints}

We assume that all background fields are constant which enables us to calculate Poisson brackets. Using standard Poisson algebra, the consistency procedure for
$\gamma_\mu^{(0)}$ produces an infinite set of constraints
\begin{equation}
\gamma_\mu^{(n)}\equiv\{H_c,\gamma_\mu^{(n-1)}\}\quad
(n=1,2,3,\dots)\, ,
\end{equation}
which can be rewritten at $\sigma=0$ in the compact
$\sigma$-dependent form
\begin{eqnarray}\label{eq:10}
\Gamma_\mu(\sigma)\equiv \sum_{n=0}^\infty
\frac{\sigma^n}{n!}\gamma_\mu^{(n)}|_0=\Pi_{+ \mu}{}^\nu I_{-
\nu}(\sigma)+\Pi_{- \mu}{}^\nu I_{+
\nu}(-\sigma)+\pi_{\alpha}(-\sigma)\Psi^{\alpha}_{\mu}+\bar\Psi^{\alpha}_\mu
\bar\pi_{\alpha}(\sigma)\, .
\end{eqnarray}

From conditions $(\theta^\alpha-\bar\theta^{\alpha})\big |_0=0$
and $(\pi_\alpha-\bar\pi_{\alpha})\big |_0=0$, we get
\begin{eqnarray}\label{eq:11}
\Gamma^\alpha(\sigma)=
\Theta^\alpha(\sigma)-\bar\Theta^{\alpha}(\sigma)\, ,\quad
\Gamma_{\alpha}^\pi(\sigma)\equiv
\Pi_\alpha(\sigma)-\bar\Pi_\alpha(\sigma)\, ,
\end{eqnarray}
where the right-hand side functions are defined as
\begin{eqnarray}\label{eq:Fialfa9}
\Theta^\alpha(\sigma)=\theta^\alpha(-\sigma)-\Psi^\alpha_\mu
\tilde q^\mu(\sigma)-\frac{1}{2\kappa}F^{\alpha\beta}\int_0^\sigma
d\sigma_1 P_s \bar\pi_{\beta} +\frac{1}{2\kappa}G^{\mu\nu}
\Psi^\alpha_\mu   \int_0^\sigma d\sigma_1 P_s (I_{+ \nu}+I_{-
\nu}),\nonumber\\
\end{eqnarray}
\begin{eqnarray}\label{eq:barFialfa9}
\bar\Theta^{\alpha}(\sigma)=\bar\theta^{\alpha}(\sigma)+\bar\Psi^{\alpha}_\mu
\tilde q^\mu(\sigma)+\frac{1}{2\kappa} F^{\beta\alpha}
\int_0^\sigma d\sigma_1 P_s \pi_{\beta} \,\,
+\frac{1}{2\kappa}G^{\mu\nu} \bar\Psi^{\alpha}_\mu \int_0^\sigma
d\sigma_1 P_s (I_{+ \nu}+I_{- \nu})\, ,\nonumber\\
\end{eqnarray}
\begin{equation}\label{eq:Pialfa9}
\Pi_\alpha(\sigma)=\pi_\alpha(-\sigma)\, ,\quad
\bar\Pi_{\bar\alpha}(\sigma)=\bar\pi_{\bar\alpha}(\sigma)\, .
\end{equation}

Similarly, for $D5$-brane boundary conditions (\ref{eq:gamami}), (\ref{eq:bcf1}) and (\ref{eq:bcf2}) we get
\begin{eqnarray}\label{eq:D5gu1}
\Gamma_i(\sigma)&=&\Pi_{+ i}{}^j I_{-
j}(\sigma)+\Pi_{- i}{}^j I_{+ j}(-\sigma)\nonumber \\
&+&2\Re
\left[\pi_{\alpha_1}(-\sigma)\Psi^{\alpha_1}_{i}+\pi_{\alpha_2}(-\sigma)\Psi^{\alpha_2}_{i}+\bar\Psi^{\alpha_1}_i
\bar\pi_{\alpha_1}(\sigma)+\bar\Psi^{\alpha_2}_i
\bar\pi_{\alpha_2}(\sigma)\right]\, ,
\end{eqnarray}
\begin{eqnarray}
\Gamma^{\alpha_1}(\sigma)=
\Theta^{\alpha_1}(\sigma)-\bar\Theta^{\alpha_1}(\sigma)\, ,\quad
\Gamma^{\alpha_2}(\sigma)=
\Theta^{\alpha_2}(\sigma)+\bar\Theta^{\alpha_2}(\sigma)\, ,
\end{eqnarray}
\begin{eqnarray}\label{eq:D5gu3}
\Gamma_{\alpha_1}^\pi(\sigma)=\pi_{\alpha_1}(-\sigma)-\bar\pi_{\alpha_1}(\sigma)\,
,\quad
\Gamma_{\alpha_2}^\pi(\sigma)=\pi_{\alpha_2}(-\sigma)+\bar\pi_{\alpha_2}(\sigma)\,
,
\end{eqnarray}
where right-hand side variables are defined as
\begin{eqnarray}
&{}&\Theta^{\alpha_1}(\sigma)=\theta^{\alpha_1}(-\sigma)-\Psi^{\alpha_1}_i
\tilde q^i(\sigma)-\frac{1}{2\kappa}
f^{\alpha_1\beta_{1}}_{11}\int_0^\sigma d\sigma_1 P_s
\bar\pi_{\beta_{1}}-\frac{1}{2\kappa}
f^{\alpha_1\beta_{1}}_{14}\int_0^\sigma d\sigma_1 P_s
\bar\pi_{\beta_{1}}^*\nonumber \\&+&\frac{1}{2\kappa}f_{12}^{\alpha_1\beta_2}\int_0^\sigma d\sigma_1 P_s \bar\pi_{\beta_2}+\frac{1}{2\kappa}f_{13}^{\alpha_1\beta_2}\int_0^\sigma d\sigma_1 P_s \bar\pi_{\beta_2}^*+\frac{1}{2\kappa}G^{ij}
\Psi^{\alpha_1}_i   \int_0^\sigma d\sigma_1 P_s (I_{+ j}+I_{- j})
,\nonumber \\ \label{eq:Fialfa}
\end{eqnarray}
\begin{eqnarray}
&{}&\bar\Theta^{\alpha_1}(\sigma)=\bar\theta^{\alpha_1}(\sigma)+\bar\Psi^{\alpha_1}_i
\tilde q^i(\sigma)+\frac{1}{2\kappa}f^{\beta_{1}\alpha_1}_{11}
\int_0^\sigma d\sigma_1 P_s
\pi_{\beta_{1}}+\frac{1}{2\kappa}f^{*\beta_{1}\alpha_1}_{14}
\int_0^\sigma d\sigma_1
P_s \pi_{\beta_{1}}^* \,\,\nonumber \\
&+&\frac{1}{2\kappa}f_{21}^{\beta_2\alpha_1}\int_0^\sigma d\sigma_1 P_s \pi_{\beta_2}+\frac{1}{2\kappa}f_{24}^{*\beta_2\alpha_1}\int_0^\sigma d\sigma_1 P_s \pi_{\beta_2}^*+\frac{1}{2\kappa}G^{ij} \bar\Psi^{\alpha_1}_i \int_0^\sigma
d\sigma_1 P_s (I_{+ j}+I_{- j})\, .\nonumber \\ \label{eq:barFialfa}
\end{eqnarray}
The expression for $\Theta^{\alpha_2}$ can be obtained from the
expression for $\Theta^{\alpha_1}$ using substitution
$\theta^{\alpha_1}\rightarrow \theta^{\alpha_2}$,
$\pi_{\alpha_1}\leftrightarrow \pi_{\alpha_2}$,
$\Psi^{\alpha_1}_i\rightarrow \Psi^{\alpha_2}_i$,
$f_{11}\rightarrow -f_{22}$, $f_{14}\rightarrow -f_{23}$,
$f_{12}\rightarrow -f_{21}$ and $f_{13}\leftrightarrow-f_{24}$. We
obtain the expression for $\bar\Theta^{\alpha_2}$ from
$\bar\Theta^{\alpha_1}$ using similar transition rules (fermionic
variables and background fields have bars).

We introduced variables, even and odd
under world-sheet parity transformation $\Omega:\sigma\to -\sigma$. For
bosonic variables we use standard notation \cite{BNBS}
\begin{equation}\label{eq:bv1}
q^\mu(\sigma)= P_s x^\mu(\sigma)\, ,\quad \tilde q^\mu(\sigma)=
P_a x^\mu(\sigma)\, ,\nonumber
\end{equation}
\begin{equation}\label{eq:bv2}
p_\mu(\sigma)= P_s \pi_\mu(\sigma)\, ,\quad \tilde p_\mu(\sigma)=
P_a \pi_\mu(\sigma)\, ,
\end{equation}
while for fermionic ones we explicitly use
the projectors on $\Omega$ even and odd parts
\begin{equation}\label{eq:PsPa}
P_s=\frac{1}{2}(1+\Omega)\, ,\quad P_a=\frac{1}{2}(1-\Omega)\, .
\end{equation}

For all constraints we apply the consistency procedure at $\sigma
=\pi$ and obtain similar expressions, where all variables
depending on $-\sigma$ are replaced by the same variables
depending on $2\pi-\sigma$. That set of constraints is solved by
$2\pi$ periodicity of all canonical variables as in
Ref.\cite{BNBS}.

\subsection{Classification of constraints}

Let us denote all constraints with $\Gamma_A=(\Gamma_\mu\,
,\Gamma^\alpha\, ,\Gamma_\alpha^\pi)$. From
\begin{equation}
\left\lbrace H_c\, ,\Gamma_A\right\rbrace=\Gamma_A'\approx0\, ,
\end{equation}
it follows that all constraints weakly commute with canonical
Hamiltonian, so there are no more constraints in the theory and
the consistency procedure is completed.

For practical reasons we will separate the constraints $\Gamma_A$ in two sets: the zero modes $(\theta^\alpha-\bar\theta^\alpha)|_0$ and the rest ${}^\star\Gamma_A=(\Gamma_\mu\, ,\Gamma'^\alpha\,
,\Gamma^\pi_\alpha)$. The reason for this separation is that Poisson brackets
of constraints ${}^\star\Gamma_A$ close on $\delta'$ function
while those with $\Gamma_A$, close on $\delta$,
$\delta'$ or step function. 

First we will classify
${}^\star{}\Gamma_A$. The algebra of the constraints ${}^\star\Gamma_A$ has the form
\begin{equation}\label{eq:AB}
\left\lbrace {}^\star\Gamma_A\,
,{}^\star\Gamma_B\right\rbrace=M_{AB}\delta'\, ,
\end{equation}
where the supermatrix $M_{AB}$ is given by the expression
\begin{equation}\label{eq:MAB}
M_{AB}=\left(
\begin{array}{cc}
(M_1)_{\mu\nu} & (M_2)_\mu{}^\gamma{}_\delta\\
(M_3)^\alpha{}_\beta{}_\nu & (M_4)^\alpha{}_\beta{}^\gamma
{}_\delta
\end{array}\right)=\left(
\begin{array}{c|cc}
-\kappa G^{eff}_{\mu\nu} & -2(\Psi_{eff})^\gamma_\mu & 0\\ \hline
-2(\Psi_{eff})^\alpha_\nu & \frac{1}{\kappa}F^{\alpha\gamma}_{eff} & -2\delta^\alpha{}_\delta \\
0 & -2\delta_\beta{}^\gamma & 0
\end{array}\right)\, .
\end{equation}
Here we introduced effective background fields
\begin{eqnarray}\label{eq:effbackground9}
G^{eff}_{\mu\nu}&=&G_{\mu\nu}-4B_{\mu\rho}G^{\rho\lambda}B_{\lambda\nu}\,
,\quad
(\Psi_{eff})^\alpha_{\mu}=\frac{1}{2}\Psi^\alpha_{+\mu}+B_{\mu\rho}G^{\rho\nu}
\Psi^\alpha_{-\nu}\, ,\nonumber \\
F_{eff}^{\alpha\beta}&=&F_{a}^{\alpha\beta}-\Psi^\alpha_{-\mu}G^{\mu\nu}\Psi^\beta_{-\nu}\,
,
\end{eqnarray}
and useful notation
\begin{equation}\label{eq:saF9}
\Psi^\alpha_{\pm \mu}=\Psi^\alpha_\mu\pm\bar\Psi^{\alpha}_\mu\,
,\quad
F_s^{\alpha\beta}=\frac{1}{2}(F^{\alpha\beta}+F^{\beta\alpha})\,
,\quad
F_a^{\alpha\beta}=\frac{1}{2}(F^{\alpha\beta}-F^{\beta\alpha})\, .
\end{equation}
Following \cite{SW} we will refer to the fields appearing in
matrix $M_{AB}$ as the \textit{open string} background fields.
This is supersymmetric generalization of Seiberg and Witten open
string metric, $G_{\mu\nu}^{eff}$, because all effective fields
contain bilinear combinations of $\Omega$ odd fields.

When $D5$-brane is embedded in ten dimensional space-time, the
boundary conditions ${}^\star\Gamma_{A}=(\Gamma_i\,
,\Gamma'^{\alpha_1}\, ,\Gamma'^{\alpha_2}\,
,\Gamma_{\alpha_1}^\pi\, ,\Gamma_{\alpha_2}^\pi)$ satisfy the
algebra (\ref{eq:AB}), where the supermatrix $M_{AB}$ is given by
the expression
\begin{eqnarray}\label{eq:algebrav}
M_{AB}=\left(
\begin{array}{c|ccccc}
-\kappa G_{ij}^{eff} & -2 (\Psi^{eff})_i{}^{\gamma_1} &
-2(\Psi^{eff})_i{}^{\gamma_2} & 0 & 0\\ \hline
-2(\Psi^{eff})^{\alpha_1}{}_j & \frac{1}{\kappa}(f_{11}^{eff})^{\alpha_1\gamma_{1}} & \frac{1}{\kappa}(f_{12}^{eff})^{\alpha_1\gamma_{2}} & -2\delta^{\alpha_1}{}_{\delta_{1}} & 0 \\
-2(\Psi^{eff})^{\alpha_2}{}_j & -\frac{1}{\kappa}(f_{12}^{eff})^{\alpha_2\gamma_{1}} & \frac{1}{\kappa}(f_{22}^{eff})^{\alpha_2\gamma_{2}} & 0 & -2\delta^{\alpha_2}{}_{\delta_{2}} \\
0 & -2\delta_{\beta_1}{}^{\gamma_{1}} & 0 & 0 & 0\\
0 & 0 & -2\delta_{\beta_2}{}^{\gamma_{2}} & 0 & 0
\end{array}\right)\, .
\end{eqnarray}
The open string background fields are defined as
\begin{eqnarray}\label{eq:effbackground}
G^{eff}_{ij}&=&G_{ij}-4B_{i k}G^{kl}B_{l j}\,
,\nonumber
\\(\Psi_{eff})^{\alpha_1}_{i}&=&\frac{1}{2}\Psi^{\alpha_1}_{+ i}+B_{ik}G^{kj}
\Psi^{\alpha_1}_{- j}\, ,\quad
(\Psi_{eff})^{\alpha_2}_{i}=\frac{1}{2}\Psi^{\alpha_2}_{-
i}+B_{ik}G^{kj} \Psi^{\alpha_2}_{+ j}\, ,\nonumber \\
(f^{eff}_{11})^{\alpha_1\beta_1}&=&(f_{11}^a)^{\alpha_1\beta_1}-\Psi^{\alpha_1}_{-i}G^{ij}\Psi^{\beta_1}_{-j}\,
,\quad
(f^{eff}_{22})^{\alpha_2\beta_2}=(f^a_{22})^{\alpha_2\beta_2}-\Psi^{\alpha_2}_{+i}G^{ij}\Psi^{\beta_2}_{+j}\,
,\nonumber \\
(f_{12}^{eff})^{\alpha_1\beta_2}&=&\frac{1}{2}\left(f_{12}^{\alpha_1\beta_2}-f_{21}^{\beta_2\alpha_1}
\right) - \Psi^{\alpha_1}_{- i} G^{ij}\Psi^{\beta_2}_{+j}\, .
\end{eqnarray}

From the definition of superdeterminant
\begin{equation}
s\det M_{AB}=\frac{\det(M_1-M_2 M_4^{-1}M_3)}{\det M_4}\, ,
\end{equation}
and using the fact that
\begin{equation}
M_2 M_4^{-1}M_3=0\, ,\quad \det M_4=const\, ,
\end{equation}
we obtain from (\ref{eq:MAB})
\begin{equation}\label{eq:detM}
s\det M_{AB}\thicksim \det G^{eff}\, .
\end{equation}
Because we assume that effective metric $G^{eff}$ is
nonsingular, we conclude that all constraints ${}^\star\Gamma_A$ are of the second class. It is easy to check that zero modes, $(\theta^\alpha-\bar\theta^\alpha)|_0$, are also of the second class, and consequently, all constraints originating from
boundary conditions, $\Gamma_A$, are of the second class. Note that the condition $s\det M_{AB}\neq0$ is
exactly the same condition as in the bosonic case \cite{BNBS}.

\section{Solution of the constraints}
\setcounter{equation}{0}

Instead to calculate Dirac brackets we prefer to explicitly solve
second class constraints originating from boundary conditions.
From $\Gamma_\mu=0$, $\Gamma^{\alpha}=0$ and
$\Gamma_{\alpha}^\pi=0$, we obtain
\begin{equation}\label{eq:resenjex9}
x^{\mu}(\sigma)=q^\mu-2\Theta^{\mu\nu}\int_0^\sigma d\sigma_1
p_\nu+2\Theta^{\mu\alpha}\int_0^\sigma d\sigma_1 p_{\alpha}\,
,\quad \pi_\mu=p_\mu\, ,
\end{equation}
\begin{eqnarray}\label{eq:resenje19}
\theta^\alpha(\sigma)=\Phi^\alpha(\sigma)+\frac{1}{2}\tilde\xi^{\alpha}\,
,\quad \pi_\alpha=p_\alpha+\tilde p_\alpha\, ,
\end{eqnarray}
\begin{eqnarray}\label{eq:resenje29}
\bar\theta^\alpha(\sigma)=\Phi^\alpha(\sigma)-\frac{1}{2}\tilde\xi^{\alpha}\,
,\quad \bar\pi_{\alpha}=p_\alpha-\tilde p_\alpha\, ,
\end{eqnarray}
where
\begin{equation}\label{eq:fialfa}
\Phi^\alpha(\sigma)=\frac{1}{2}\xi^\alpha-\Theta^{\mu\alpha}\int_0^\sigma
d\sigma_1 p_\mu -\Theta^{\alpha\beta}\int_0^\sigma d\sigma_1
p_\beta\, ,
\end{equation}
\begin{eqnarray}\label{eq:resenjepi9}
&{}&\frac{1}{2}\xi^\alpha\equiv
P_s\theta^\alpha=P_s\bar\theta^{\alpha}\, ,\quad
\tilde\xi^{\alpha}\equiv P_a(\theta^\alpha-\bar\theta^{\alpha})\,
,\nonumber\\ &{}& p_\alpha\equiv P_s\pi_\alpha=P_s\bar\pi_{\alpha}\, ,\quad
\tilde p_\alpha\equiv P_a\pi_\alpha=-P_a\bar\pi_{\alpha}   \, ,
\end{eqnarray}
and
\begin{equation}\label{eq:teta19}
\Theta^{\mu\nu}=-\frac{1}{\kappa}(G_{eff}^{-1}BG^{-1})^{\mu\nu}\,
,\quad
\Theta^{\mu\alpha}=2\Theta^{\mu\nu}(\Psi_{eff})^\alpha_{\nu}-\frac{1}{2\kappa}G^{\mu\nu}\psi^\alpha_{-\nu}\,
,
\end{equation}
\begin{eqnarray}\label{eq:teta29}
\Theta^{\alpha\beta}&=&\frac{1}{2\kappa}F_s^{\alpha\beta}+4(\Psi_{eff})^\alpha_{\mu}
\Theta^{\mu\nu}(\Psi_{eff})^\beta_{\nu}-\frac{1}{\kappa}\Psi^\alpha_{-\mu}(G^{-1}BG^{-1})^{\mu\nu}\Psi^\beta_{-\nu}\nonumber \\ &+&\frac{G^{\mu\nu}}{\kappa}\left[ \Psi^\alpha_{-\mu}(\Psi_{eff})^\beta_{\nu}+\Psi^\beta_{-\mu}(\Psi_{eff})^\alpha_{\nu}
\right] \,
.
\end{eqnarray}

Using $\sigma$-dependent form of boundary conditions (\ref{eq:D5gu1})-(\ref{eq:D5gu3}), we get $D5$-brane variables in terms of effective
ones
\begin{equation}\label{eq:resenjex}
x^{i}(\sigma)=q^i-2\Theta^{ij}\int_0^\sigma d\sigma_1
p_j+4\Re\left(\Theta^{i\alpha_1}\int_0^\sigma d\sigma_1 p_{\alpha_1}+\Theta^{i\alpha_2}\int_0^\sigma d\sigma_1 p_{\alpha_2}\right)\,
,\quad \pi_i=p_i\, ,
\end{equation}
\begin{eqnarray}\label{eq:resenje1}
\theta^{\alpha_1}(\sigma)&=&\Phi^{\alpha_1}(\sigma)+\frac{1}{2}\tilde\xi^{\alpha_1}(\sigma)\,
,\quad \pi_{\alpha_1}=p_{\alpha_1}+\tilde p_{\alpha_1}\, ,
\end{eqnarray}
\begin{eqnarray}\label{eq:resenje12}
\theta^{\alpha_2}(\sigma)=\Phi^{\alpha_2}(\sigma)+\frac{1}{2}\tilde\xi^{\alpha_2}(\sigma)\,
,\quad \pi_{\alpha_2}=p_{\alpha_2}+\tilde p_{\alpha_2}\, ,
\end{eqnarray}
\begin{eqnarray}
\bar\theta^{\alpha_1}(\sigma)=\Phi^{\alpha_1}(\sigma)-\frac{1}{2}\tilde
\xi^{\alpha_1}(\sigma)\, ,\quad
\bar\pi_{\alpha_1}=p_{\alpha_1}-\tilde p_{\alpha_1}\, ,
\end{eqnarray}
\begin{eqnarray}\label{eq:resenje22}
\bar\theta^{\alpha_2}(\sigma)=-\Phi^{\alpha_2}(\sigma)+\frac{1}{2}\tilde\xi^{\alpha_2}(\sigma)\,
, \quad \bar\pi_{\alpha_2}=-p_{\alpha_2}+\tilde p_{\alpha_2}\, ,
\end{eqnarray}
where
\begin{eqnarray}
\Phi^{\alpha_1}(\sigma)&=&\frac{1}{2}\xi^{\alpha_1}-\Theta^{i\alpha_1}\int_0^\sigma
d\sigma_1 p_i -\Theta^{\alpha_1\beta_{1}}\int_0^\sigma d\sigma_1
p_{\beta_{1}}-\Theta^{\alpha_1\beta_{2}}\int_0^\sigma d\sigma_1
p_{\beta_{2}}\nonumber
\\&-&{}^\star\Theta^{\alpha_1\beta_1}\int_0^\sigma d\sigma_1
p^*_{\beta_{1}}-{}^\star\Theta^{\alpha_1\beta_2}\int_0^\sigma
d\sigma_1 p^*_{\beta_{2}}\, ,
\end{eqnarray}
\begin{eqnarray}
\Phi^{\alpha_2}(\sigma)&=&\frac{1}{2}\xi^{\alpha_2}-\Theta^{i\alpha_2}\int_0^\sigma
d\sigma_1 p_i-\Theta^{\alpha_2\beta_1}\int_0^\sigma d\sigma_1
p_{\beta_{1}}-\Theta^{\alpha_2\beta_2}\int_0^\sigma d\sigma_1
p_{\beta_{2}}\nonumber \\ &-&{}^\star
\Theta^{\alpha_2\beta_1}\int_0^\sigma d\sigma_1
p^*_{\beta_{1}}-{}^\star \Theta^{\alpha_2\beta_2}\int_0^\sigma
d\sigma_1 p^*_{\beta_{2}}\, ,
\end{eqnarray}
and the coefficients multiplying momenta are of the form
\begin{eqnarray}\label{eq:teta1}
\Theta^{ij}&=&-\frac{1}{\kappa}(G_{eff}^{-1}BG^{-1})^{ij}\,
, \\
\Theta^{i \alpha_1}&=&2\Theta^{ij}(\Psi_{eff})^{\alpha_1}_j-\frac{1}{2\kappa}G^{ij}\Psi^{\alpha_1}_{- j}\,
,\quad \Theta^{i \alpha_2}=2\Theta^{ij}(\Psi_{eff})^{\alpha_2}_j-\frac{1}{2\kappa}G^{ij}\Psi^{\alpha_2}_{+ j}\, ,
\end{eqnarray}
\begin{eqnarray}\label{eq:teta2}
\Theta^{\alpha_1\beta_{1}}&=&\frac{1}{2\kappa}(f^s_{11})^{\alpha_1 \beta_{1}}+4(\Psi_{eff})^{\alpha_1}_i
\Theta^{ij}(\Psi_{eff})^{\beta_{1}}_j-\frac{1}{\kappa}\Psi^{\alpha_1}_{- i}(G^{-1}BG^{-1})^{ij}\Psi^{\beta_{1}}_{- j}\nonumber \\ &+&\frac{G^{ij}}{\kappa}\left[ \Psi^{\alpha_1}_{- i}(\Psi_{eff})^{\beta_{1}}_j+\Psi^{\beta_{1}}_{- i}(\Psi_{eff})^{\alpha_{1}}_j
\right] \,
,
\end{eqnarray}
\begin{eqnarray}\label{eq:teta2222}
\Theta^{\alpha_1\beta_{2}}&=&\Theta^{\beta_2\alpha_1}=\frac{1}{4\kappa}(f_{12}^{\alpha_1\beta_2}+f_{21}^{\beta_2\alpha_1})+4(\Psi_{eff})^{\alpha_1}_i
\Theta^{ij}(\Psi_{eff})^{\beta_{2}}_j\nonumber \\ &-&\frac{1}{\kappa}\Psi^{\alpha_1}_{- i}(G^{-1}BG^{-1})^{ij}\Psi^{\beta_{2}}_{+ j}+\frac{G^{ij}}{\kappa}\left[ \Psi^{\alpha_1}_{- i}(\Psi_{eff})^{\beta_{2}}_j+\Psi^{\beta_{2}}_{+ i}(\Psi_{eff})^{\alpha_{1}}_j
\right] \,
,
\end{eqnarray}
\begin{eqnarray}\label{eq:teta22}
{}^\star\Theta^{\alpha_1\beta_{1}}&=&\frac{1}{4\kappa}(f_{14}^{\alpha_1\beta_1}+f_{14}^{*\beta_1\alpha_1})+4(\Psi_{eff})^{\alpha_1}_i
\Theta^{ij}(\Psi_{eff})^{*\beta_{1}}_j-\frac{1}{\kappa}\Psi^{\alpha_1}_{- i}(G^{-1}BG^{-1})^{ij}\Psi^{*\beta_{1}}_{- j}\nonumber \\ &+&\frac{G^{ij}}{\kappa}\left[ \Psi^{\alpha_1}_{- i}(\Psi_{eff})^{*\beta_{1}}_j+\Psi^{*\beta_{1}}_{- i}(\Psi_{eff})^{\alpha_{1}}_j
\right] \,
,
\end{eqnarray}
\begin{eqnarray}\label{eq:tetaz12}
{}^\star\Theta^{\alpha_1\beta_{2}}&=&{}^\star \Theta^{\beta_2\alpha_1}=\frac{1}{4\kappa}(f_{13}^{\alpha_1\beta_2}+f_{24}^{*\beta_2\alpha_1})+4(\Psi_{eff})^{\alpha_1}_i
\Theta^{ij}(\Psi_{eff})^{*\beta_{2}}_j\nonumber \\&-&\frac{1}{\kappa}\Psi^{\alpha_1}_{- i}(G^{-1}BG^{-1})^{ij}\Psi^{*\beta_{2}}_{+ j}+\frac{G^{ij}}{\kappa}\left[ \Psi^{\alpha_1}_{- i}(\Psi_{eff})^{*\beta_{2}}_j+\Psi^{*\beta_{2}}_{+ i}(\Psi_{eff})^{\alpha_{1}}_j
\right] \, .
\end{eqnarray}
The coefficient $\Theta^{\alpha_2\beta_2}$ can be obtained from
$\Theta^{\alpha_1\beta_1}$ after substitution $f_{11}^s\to
f_{22}^s$, $(\Psi_{eff})^{\alpha_1}_i\to
(\Psi_{eff})^{\alpha_2}_i$ and $\Psi^{\alpha_1}_{-i}\to
\Psi^{\alpha_2}_{+i}$, while ${}^\star\Theta^{\alpha_2\beta_2}$
follows from ${}^\star\Theta^{\alpha_1\beta_1}$ after substitution
$f_{14}\to f_{23}$, $(\Psi_{eff})^{\alpha_1}_i\to
(\Psi_{eff})^{\alpha_2}_i$ and $\Psi^{\alpha_1}_{-i}\to
\Psi^{\alpha_2}_{+i}$.

\section{Noncommutativity of $Dp$-brane world-volume}
\setcounter{equation}{0}

Using the solutions of boundary conditions we will show that
Poisson brackets of initial $Dp$-brane variables are nonzero.

\subsection{$D9$-brane}

From basic Poisson algebra
\begin{equation}
\{x^\mu(\sigma), \pi_\nu(\bar\sigma)\}=\delta^\mu{}_\nu
\delta(\sigma-\bar\sigma)\, ,
\end{equation}
and definitions (\ref{eq:bv1})-(\ref{eq:bv2}) we obtain
\begin{equation}\label{eq:pz0}
\{q^\mu(\sigma)\, ,p_\nu(\bar\sigma)\}=\delta^\mu{}_\nu
\delta_s(\sigma\, ,\bar\sigma)\, ,\quad \{\tilde q^\mu(\sigma)\,
,\tilde p_\nu(\bar\sigma)\}=\delta^\mu{}_\nu \delta_a(\sigma\,
,\bar\sigma)\, ,
\end{equation}
where
\begin{equation}
\delta_s(\sigma,\bar\sigma)=\frac{1}{2}\left[
\delta(\sigma-\bar\sigma)+\delta(\sigma+\bar\sigma)\right]\,
,\quad \delta_a(\sigma,\bar\sigma)=\frac{1}{2}\left[
\delta(\sigma-\bar\sigma)-\delta(\sigma+\bar\sigma)\right]\, ,
\end{equation}
are symmetric and antisymmetric delta functions, respectively.
Using basic Poisson algebra of fermionic variables
\begin{equation}
\{\theta^\alpha(\sigma),\pi_\beta(\bar\sigma)\}=\{\bar\theta^\alpha(\sigma),\bar\pi_\beta(\bar\sigma)\}=-\delta^\alpha{}_\beta\delta(\sigma-\bar\sigma)\,
,\quad
\end{equation}
we have
\begin{equation}
\left\lbrace
P_s\theta^{\alpha}(\sigma)+P_s\bar\theta^{\alpha}(\sigma)\,
,P_s \pi_{\beta}(\bar\sigma)+P_s\bar\pi_{\beta}(\bar\sigma)\right\rbrace=
-2\delta^{\alpha}{}_{\beta} \delta_s(\sigma\, ,\bar\sigma)\, ,
\end{equation}
which gives
\begin{equation}
\left\lbrace \xi^{\alpha}(\sigma)\,
,p_{\beta}(\bar\sigma)\right\rbrace =-\delta^{\alpha}{}_{\beta}
\delta_s(\sigma\, ,\bar\sigma)\, .
\end{equation}
Similarly we obtain
\begin{equation}
\left\lbrace \tilde\xi^{\alpha}(\sigma)\, ,\tilde
p_{\beta}(\bar\sigma)\right\rbrace
=-\delta^{\alpha}{}_{\beta}\delta_a(\sigma\, ,\bar\sigma)\, .
\end{equation}
Therefore, the momenta $p_\mu$, $\tilde p_\mu$, $p_{\alpha}$ and $\tilde p_{\alpha}$ are
canonically conjugated to the coordinates $q^\mu$, $\tilde q^\mu$,
$\xi^{\alpha}$ and $\tilde \xi^{\alpha}$, respectively.

Using  the solutions of constraints
(\ref{eq:resenjex9})-(\ref{eq:resenje29}), we get the
noncommutativity relations
\begin{equation}\label{eq:nc91}
\{x^\mu(\sigma)\, ,x^\nu(\bar\sigma)\}=2
\Theta^{\mu\nu}\theta(\sigma+\bar\sigma)\, ,
\end{equation}
\begin{equation}\label{eq:nc92}
\{x^\mu(\sigma)\,
,\theta^\alpha(\bar\sigma)\}=-\Theta^{\mu\alpha}\theta(\sigma+\bar\sigma)\,
,\quad \{\theta^\alpha(\sigma)\,
,\bar\theta^{\beta}(\bar\sigma)\}=\frac{1}{2}\Theta^{\alpha\beta}\theta(\sigma+\bar\sigma)\,
,
\end{equation}
where
\begin{equation}\label{eq:fdelt}
\theta(x)=\left\{\begin{array}{ll}
0 & \textrm{if $x=0$}\\
1/2 & \textrm{if $0<x<2\pi$}\, .\\
1 & \textrm{if $x=2\pi$} \end{array}\right .
\end{equation}
After introducing center of mass variables
\begin{equation}
A(\sigma)=A_{cm}+\mathcal A(\sigma)\, ,\quad A_{cm}=\frac{1}{\pi}\int_0^\pi d\sigma A(\sigma)\, ,
\end{equation}
where $A(\sigma)$ is arbitrary variable, we obtain
\begin{equation}\label{eq:ncx9}
\{x^\mu(\sigma)\, ,x^\nu(\bar\sigma)\}=
\Theta^{\mu\nu}\Delta(\sigma+\bar\sigma)\, ,
\end{equation}
\begin{equation}\label{eq:ncteta9}
\{x^\mu(\sigma)\,
,\theta^\alpha(\bar\sigma)\}=-\frac{1}{2}\Theta^{\mu\alpha}\Delta(\sigma+\bar\sigma)\,
,\quad \{\theta^\alpha(\sigma)\,
,\bar\theta^{\beta}(\bar\sigma)\}=\frac{1}{4}\Theta^{\alpha\beta}\Delta(\sigma+\bar\sigma)\,
.
\end{equation}
The function $\Delta(\sigma+\bar\sigma)$ is nonzero only at string
endpoints
\begin{equation}\label{eq:DELTA}
\Delta(x)=2\theta(x)-1=\left\{\begin{array}{ll}
-1 & \textrm{if $x=0$}\\
0 & \textrm{if $0<x<2\pi$}\, ,\\
1 & \textrm{if $x=2\pi$} \end{array}\right .
\end{equation}
and we conclude that interior of the string is commutative, while
string endpoints are noncommutative.

\subsection{$D5$-brane}

Applying the same procedure as in the case of $D9$-brane, we get
$D5$-brane noncommutativity relations
\begin{equation}\label{eq:ncx}
\{x^i(\sigma)\, ,x^j(\bar\sigma)\}=
\Theta^{ij}\Delta(\sigma+\bar\sigma)\, ,
\end{equation}
\begin{eqnarray}
\{x^i(\sigma)\,
,\theta^{\alpha_1}(\bar\sigma)\}=-\frac{1}{2}\Theta^{i\alpha_1}\Delta(\sigma+\bar\sigma)\,
&,&\; \{x^i(\sigma)\,
,\theta^{\alpha_2}(\bar\sigma)\}=-\frac{1}{2}\Theta^{i\alpha_2}\Delta(\sigma+\bar\sigma)\,
,
\end{eqnarray}
\begin{eqnarray}
\{\theta^{\alpha_1}(\sigma)\,
,\bar\theta^{\beta_{1}}(\bar\sigma)\}=\frac{1}{4}\Theta^{\alpha_1\beta_{1}}\Delta(\sigma+\bar\sigma)\,
,\quad \{\theta^{\alpha_2}(\sigma)\,
,\bar\theta^{\beta_{2}}(\bar\sigma)\}=-\frac{1}{4}\Theta^{\alpha_2\beta_{2}}\Delta(\sigma+\bar\sigma)\,
,
\end{eqnarray}
\begin{eqnarray}\label{eq:ncteta}
\{\theta^{\alpha_1}(\sigma)\,
,\bar\theta^{\beta_{2}}(\bar\sigma)\}=-\frac{1}{4}\Theta^{\alpha_1\beta_2}\Delta(\sigma+\bar\sigma)\, .
\end{eqnarray}

The parameters
multiplying complex conjugated momenta denoted by star are absent in
noncommutativity relations, because the solutions for initial
fermionic coordinates do not depend on complex conjugated
effective coordinates.

On the solutions of the boundary conditions original string
variables depend both on effective coordinates and effective
momenta, and that is a source of noncommutativity. In the
supersymmetric case the presence of $\Omega$ odd fields $B_{\mu\nu}$,
$\Psi^{\alpha}_{-\mu}$ and
$F_{s}^{\alpha\beta}$
leads to noncommutativity of the supercoordinates. Nontrivial
$B_{\mu\nu}$ leads to nonzero of all noncommutative parameters,
$\Theta^{\mu\nu}$, $\Theta^{\mu\alpha}$ and
$\Theta^{\alpha\beta}$. If only $\Psi^{\alpha}_{-\mu}$ is nontrivial, we have $\Theta^{\mu\nu}=0$,
but $\Theta^{\mu\alpha}$ and
$\Theta^{\alpha\beta}$ are nonzero. Finally, if only
$F_{s}^{\alpha\beta}$ is nontrivial then $\Theta^{\mu\nu}=0$ and $\Theta^{\mu\alpha}=0$, and only $\Theta^{\alpha\beta}$ is
nonzero. The last case corresponds to the noncommutativity
relations used in \cite{voja}, where bosonic variables are commutative. This discussion is the same for $D5$-brane up to the following replacing
\begin{eqnarray}
&{}&\Psi^\alpha_{\pm \mu}=\Psi^\alpha_\mu\pm\bar\Psi^\alpha_\mu \to \Psi^\alpha_{\pm \mu}=\Psi^\alpha_\mu\mp({}^\star\Gamma\bar\Psi_\mu)^\alpha\, ,\nonumber \\ &{}& F_s^{\alpha\beta}=\frac{1}{2}(F^{\alpha\beta}+F^{\beta\alpha})\to F_s^{\alpha\beta}=-\frac{1}{2}\left[(F{}^\star\Gamma)^{\alpha\beta}+(F{}^\star\Gamma)^{\beta\alpha}\right]\, .
\end{eqnarray}

\section{Supersymmetry of noncommutativity relations}
\setcounter{equation}{0}

Here we will explicitly show that noncommutativity relations of
$D9$-brane coordinates, (\ref{eq:nc91}) and
(\ref{eq:nc92}),
are connected by $N=1$ supersymmetry transformations. Because of
the relation between $D9$ and $D5$-brane spinors \cite{BNBSNPB},
the similar relations hold for $D5$-brane supersymmetry.

The action of initial theory (\ref{eq:SB}) is invariant under
global $N=2$ supersymmetry with parameters $\epsilon$ and $\bar\epsilon$. The supersymmetry transformations of
the variables $x^\mu$, $\theta^\alpha$ and $\bar\theta^\alpha$ \cite{fabio} are
\begin{eqnarray}\label{eq:susyII}
\delta x^\mu=\bar\epsilon^{\alpha} \Gamma^\mu_{\alpha\beta} \theta^{\beta}+\epsilon^{\alpha} \Gamma^\mu_{\alpha\beta}
\bar\theta^{\beta}\, ,\quad \delta\theta^\alpha=\epsilon^\alpha\, ,\quad
\delta \bar\theta^\alpha=\bar\epsilon^\alpha\, ,
\end{eqnarray}
while the transformation rules of constant background fields are
\begin{equation}
\delta G_{\mu\nu}=\epsilon_+^\alpha\Gamma_{[\mu\,\alpha\beta}
\Psi^\beta_{+\nu]}-\epsilon_-^\alpha\Gamma_{[\mu\,\alpha\beta} \Psi^\beta_{-\nu]}\, ,\quad \delta
B_{\mu\nu}=\epsilon_+^\alpha\Gamma_{[\mu\,\alpha\beta}
\Psi^\beta_{-\nu]}-\epsilon_-^\alpha\Gamma_{[\mu\,\alpha\beta} \Psi^\beta_{+\nu]}\, ,
\end{equation}
\begin{equation}
\delta\Psi^\alpha_{+\mu}=-\frac{1}{16}\epsilon_-^\beta\Gamma_{\mu\,\beta\gamma} F_s^{\gamma\alpha}
-\frac{1}{16}\epsilon_+^\beta\Gamma_{\mu\,\beta\gamma} F_a^{\gamma\alpha}\, ,\quad
\delta\Psi^\alpha_{-\mu}=\frac{1}{16}\epsilon_+^\beta\Gamma_{\mu\,\beta\gamma} F_s^{\gamma\alpha}
+\frac{1}{16}\epsilon_-^\beta\Gamma_{\mu\,\beta\gamma} F_a^{\gamma\alpha}\, ,
\end{equation}
\begin{equation}
\delta A^{(0)}=0\, ,\quad \delta
A^{(2)}_{\mu\nu}=-\epsilon_+^\alpha\Gamma_{[\mu\,\alpha\beta}\Psi^\beta_{+\nu]}-\epsilon_-^\alpha\Gamma_{[\mu\,\alpha\beta}\Psi^\beta_{-\nu]}+A^{(0)}\delta
B_{\mu\nu}\, ,
\end{equation}
\begin{equation}
\delta
A^{(4)}_{\mu\nu\rho\sigma}=2\epsilon_+^\alpha\Gamma_{[\mu\nu\rho\,\alpha\beta}\Psi^\beta_{-\sigma]}+2\epsilon_-^\alpha\Gamma_{[\mu\nu\rho\,\alpha\beta}\Psi^\beta_{+\sigma]}+6A^{(2)}_{[\mu\nu}\delta
B_{\rho\sigma]}\, .
\end{equation}
Here we used notation
\begin{equation}
\epsilon^\alpha_{\pm}=\epsilon^\alpha\pm\bar\epsilon^\alpha=const.\,
,\quad \Gamma_{\m_1\mu_2\dots\mu_k}\equiv
\Gamma_{[\mu_1}\Gamma_{\mu_2}\dots\Gamma_{\mu_k]}\, ,
\end{equation}
and $[\;]$ in the subscripts of the fields mean antisymmetrization of space-time indices between brackets.
The potentials $A^{(0)}$ and $A^{(4)}_{\mu\nu\rho\sigma}$
correspond to the symmetric part of $F^{\alpha\beta}$
\begin{equation}
F_s^{\alpha\beta}=\frac{1}{2}(F^{\alpha\beta}+F^{\beta\alpha})\, ,
\end{equation}
and $A^{(2)}_{\mu\nu}$ to antisymmetric one
\begin{equation}
F_a^{\alpha\beta}=\frac{1}{2}(F^{\alpha\beta}-F^{\beta\alpha})\, .
\end{equation}
More about connection between two descriptions of R-R sector is given in Ref.\cite{grk} and Appendix B of Ref.\cite{BNBSNPB}.

From the solution of boundary conditions
(\ref{eq:resenje19})-(\ref{eq:resenje29}) and supersymmetry
transformations (\ref{eq:susyII}), we have
\begin{equation}
\delta\theta^\alpha(\sigma)=\delta\Phi^\alpha(\sigma)+\frac{1}{2}\delta\tilde\xi^\alpha(\sigma)=\epsilon^\alpha\,
,
\end{equation}
\begin{equation}
\delta\bar\theta^\alpha(\sigma)=\delta\Phi^\alpha(\sigma)-\frac{1}{2}\delta\tilde\xi^\alpha(\sigma)=\bar\epsilon^\alpha\,
,
\end{equation}
which gives
\begin{equation}
\delta\Phi^\alpha(\sigma)=\frac{1}{2}\epsilon_+^\alpha\, ,\quad
\delta\tilde\xi^\alpha(\sigma)=\epsilon_-^\alpha\, .
\end{equation}
From the boundary conditions (\ref{eq:gu1}), with the help of supersymmetry transformations
(\ref{eq:susyII}), it holds
\begin{equation}
\epsilon^\alpha_-=0\, .
\end{equation}

The starting $N=2$ supersymmetry transformations
(\ref{eq:susyII}), on the solution of boundary conditions, reduces
to $N=1$ supersymmetry transformations
\begin{equation}\label{eq:susyI}
\delta x^\mu(\sigma)=\epsilon^\alpha_+
\Gamma^\mu_{\alpha\beta}\Phi^\beta(\sigma)=\epsilon_+^\alpha
\Gamma^\mu_{\alpha\beta}
\theta^\beta(\sigma)-\frac{1}{2}\epsilon_+^\alpha\Gamma^\mu_{\alpha\beta}
\tilde\xi^\beta(\sigma)\, ,\quad
\delta\theta^\alpha=\delta\bar\theta^\alpha=\frac{1}{2}\epsilon_+^\alpha\,
,
\end{equation}
which gives
\begin{equation}
\delta q^\mu=\frac{1}{2}\epsilon_+^\alpha \Gamma^\mu_{\alpha\beta}
\xi^\beta\, ,\quad \delta \xi^\alpha=\epsilon_+^\alpha\, ,\quad
\delta \tilde\xi^\alpha=0\, .
\end{equation}
From Ref.\cite{GPMN} we read the supersymmetry transformations for
the momenta
\begin{equation}\label{eq:susyp}
\delta p_\alpha =
\frac{1}{2}\epsilon_+^\beta\Gamma^\mu_{\beta\alpha} p_\mu\, ,\quad
\delta p_\mu=0\, .
\end{equation}

The truncation from $N=2$ to $N=1$ supersymmetry we can realize
omitting transformations for $G_{\mu\nu}$, $\Psi_{+\mu}$ and $F_a$
\cite{fabio}. The rest fields make $N=1$ supermultiplet with
transformation rules
\begin{equation}\label{eq:susyIbfc}
\delta B_{\mu\nu}=\epsilon_+^\alpha\Gamma_{[\mu\,\alpha\beta} \Psi^\beta_{-\nu]}\, ,\quad
\delta\Psi^\alpha_{-\mu}=\frac{1}{16}\epsilon_+^\beta\Gamma_{\mu\,\beta\gamma} F_s^{\gamma\alpha}\, ,\quad
\delta F_s^{\alpha\beta}=0\, .
\end{equation}

Using $N=1$ SUSY transformations
(\ref{eq:susyI})-(\ref{eq:susyp}), we can find the supersymmetric
transformations of the coefficients $\Theta^{\mu\nu}$,
$\Theta^{\mu\alpha}$ and $\Theta^{\alpha\beta}$ multiplying the
momenta in the solutions of boundary conditions. From
\begin{eqnarray}
\delta x^{\mu}(\sigma)&=&\frac{1}{2}\epsilon_+^\alpha
\Gamma^\mu_{\alpha\beta}
\xi^\beta-2\delta\Theta^{\mu\nu}\int_0^\sigma d\sigma_1
p_\nu-2\Theta^{\mu\nu}\int_0^\sigma d\sigma_1 \delta p_\nu\,
,\\&+&2\delta\Theta^{\mu\alpha}\int_0^\sigma d\sigma_1
p_{\alpha}+2\Theta^{\mu\alpha}\int_0^\sigma d\sigma_1 \delta
p_{\alpha}=\epsilon_+^\alpha\Gamma^\mu_{\alpha\beta}\theta(\sigma)^\beta-\frac{1}{2}\epsilon_+^\alpha\Gamma^\mu_{\alpha\beta}
\tilde\xi(\sigma)^\beta\, ,\nonumber
\end{eqnarray}
\begin{equation}
\delta
\theta^{\alpha}(\sigma)=\frac{1}{2}\epsilon_+^\alpha-\delta\Theta^{\mu\alpha}\int_0^\sigma
d\sigma_1 p_\mu-\Theta^{\mu\alpha}\int_0^\sigma d\sigma_1 \delta
p_\mu -\delta\Theta^{\alpha\beta}\int_0^\sigma d\sigma_1
p_\beta-\Theta^{\alpha\beta}\int_0^\sigma d\sigma_1 \delta
p_\beta=\frac{1}{2}\epsilon_+^\alpha\, ,
\end{equation}
we obtain global $N=1$ SUSY transformations of the background
fields
\begin{equation}\label{eq:susyIbf}
\delta \Theta^{\mu\nu}=\epsilon_+^\alpha
\Gamma_{\alpha\beta}^{[\mu}\Theta^{\nu]\beta}\, ,\quad \delta
\Theta^{\mu\alpha}=-\frac{1}{2}\epsilon_+^\beta
\Gamma^\mu_{\beta\gamma} \Theta^{\gamma\alpha}\, ,\quad \delta
\Theta^{\alpha\beta}=0\, .
\end{equation}
Consequently, these fields
are components of $N=1$ supermultiplet. The coefficients, $\Theta^{\mu\nu}$, $\Theta^{\mu\alpha}$ and
$\Theta^{\alpha\beta}$, are the background fields of the
T-dual theory. This explains the fact that their SUSY transformations have the same form as
the transformations of the corresponding dual partners
$B_{\mu\nu}$, $\Psi^\alpha_{-\mu}$ and $F_s^{\alpha\beta}$
(\ref{eq:susyIbfc}).

Using $N=1$ supersymmetry transformations of SUSY coordinates (\ref{eq:susyI}) and background fields (\ref{eq:susyIbf}), we can easily prove that noncommutativity relations, (\ref{eq:nc91}) and
(\ref{eq:nc92}), are connected by supersymmetry transformations. The SUSY transformation of (\ref{eq:nc91})
\begin{equation}
\epsilon_+^\alpha \Gamma_{\alpha\beta}^{[\mu}\left\lbrace x^{\nu]},\theta^\beta\right\rbrace=-\epsilon_+^\alpha \Gamma^{[\mu}_{\alpha\beta}\Theta^{\nu]\beta}\theta(\sigma+\bar\sigma)\, ,
\end{equation}
produces the first relation in (\ref{eq:nc92}). Similarly, SUSY transformation of the first relation in (\ref{eq:nc92})
\begin{equation}
\epsilon_+^\beta \Gamma^\mu_{\beta\gamma}\left\lbrace \theta^\gamma,\theta^\alpha\right\rbrace=\frac{1}{2}\epsilon_+^\beta \Gamma^\mu_{\beta\gamma}\Theta^{\gamma\alpha}\theta(\sigma+\bar\sigma)\, ,
\end{equation}
produces the second relation in (\ref{eq:nc92}).

\section{Concluding remarks}
\setcounter{equation}{0}

In this paper we considered noncommutativity properties and related supersymmetry transformations of $D9$
and $D5$-branes in type IIB superstring theory. We used the pure
spinor formulation of the theory introduced in
Refs.\cite{berko,susyNC}.

Following \cite{kanonski,BNBSNPB,BNBS} we treated all
boundary conditions at string endpoints as canonical constraints
and checked their consistency. For nonsingular $G^{eff}$ all constraints are of the second
classand. We can solve them and obtain
the initial coordinates $x^\mu$, $\theta^{\alpha}$ and
$\bar\theta^{\alpha}$ in terms of effective ones, $q^\mu$,
$\xi^{\alpha}$ and $\tilde\xi^{\alpha}$ (momenta independent parts
of the solutions for initial supercoordinates $x^\mu$,
$\theta^{\alpha}$ and $\bar\theta^{\alpha}$) and momenta $p_\mu$
and $p_{\alpha}$ (canonically conjugated to $q^\mu$ and
$\xi^{\alpha}$).

The fact that original string variables depend both on effective
coordinates and effective momenta is a source of
noncommutativity (\ref{eq:ncx9})-(\ref{eq:ncteta9}). The solution for initial variables
do not depend on momenta $\tilde p_{\alpha}$  (canonically conjugated to $\tilde\xi^{\alpha}$). So, $\Omega$ odd
variables, denoted with tilde, do not
contribute to noncommutativity relations. Absence of the
fermionic coordinates in the solution for $x^\mu$ implies that
Poisson bracket $\left\lbrace x^\mu\, ,x^\nu\right\rbrace$ is the same
as in pure bosonic case. Similar conclusions are valid for $D5$-brane.

Noncommutativity obtained in the present paper represents a
generalization of the results from Ref.\cite{susyNC}. In special case,
when $\Psi^\alpha=\bar\Psi^\alpha_\mu$, the noncommutatativity
relations (\ref{eq:ncx9})-(\ref{eq:ncteta9}) correspond to the
relations of Ref.\cite{susyNC}.

The result of the present paper can be considered as a
supersymmetric generalization of the result obtained for bosonic
string \cite{SW}. Beside $B_{\mu\nu}$, its superpartners
$\Psi_{-\mu}^\alpha$ and $F_s^{\alpha\beta}$ are also a source of
noncommutativity. For noncommutativity of bosonic coordinates it
is necessary to have nontrivial $B_{\mu\nu}$. Noncommutativity of
bosonic and fermionic coordinates can be caused by both
$B_{\mu\nu}$ and $\Psi_{-\mu}^\alpha$, while noncommutativity of
two fermionic coordinates can be caused by all components of
noncommutative supermultiplet, $B_{\mu\nu}$, $\Psi_{-\mu}^\alpha$
and $F_s^{\alpha\beta}$.

Note that fermionic boundary conditions split $N=2$ supermultiplet
(consisting of background fields $G_{\mu\nu}$, $B_{\mu\nu}$,
$\Psi^\alpha_{+\mu}$, $\Psi^\alpha_{-\mu}$ and $F^{\alpha\beta}$)
into two $N=1$ supermultiplets. One, $\Omega$ even ($G_{\mu\nu}$,
$\Psi^\alpha_{+\mu}$, $F_a^{\alpha\beta}$), represents background
fields of type I theory, and the second one, $\Omega$ odd
($B_{\mu\nu}$, $\Psi_{-\mu}^\alpha$, $F_s^{\alpha\beta}$) is
source of noncommutativity of supercoordinates $(x^\mu\,
,\theta^\alpha)$.

\end{document}